\documentclass[reprint,
 amsmath,amssymb,
prl, 
]{revtex4-1}
\usepackage{graphicx}% Include figure files
\usepackage{dcolumn}% Align table columns on decimal point
\usepackage{bm}

\graphicspath{{./images/}}
\newcommand{\x}[0]{\mathbf{x}}
\newcommand{\dx}[0]{\mathrm{d}\x}
\DeclareMathOperator{\EX}{\mathbb{E}}

\begin{document}

\title{Determining Free Energy Differences Through Variational Morphing}% Force line breaks with \\

\author{Martin Reinhardt}

\author{Helmut Grubm\"uller}%
 \email{hgrubmu@gwdg.de}
\affiliation{%
 Max Planck Institute for Biophysical Chemistry, Am Fassberg 11, 37077 G\"ottingen, Germany
}%

\date{\today}

\begin{abstract}
Free energy calculations based on atomistic Hamiltonians and sampling are key to a first principles understanding of biomolecular processes, material properties, and macromolecular chemistry. Here, we generalize the Free Energy Perturbation method and derive non-linear Hamiltonian transformation sequences for optimal sampling accuracy that differ markedly from established linear transformations. We show that our sequences are also optimal for the Bennett Acceptance Ratio (BAR) method, and our unifying framework generalizes BAR to small sampling sizes and non-Gaussian error distributions. Simulations on a Lennard-Jones gas show that an order of magnitude less sampling is required compared to established methods.
\end{abstract}

\maketitle

Free energy calculations provide essential insights into numerous physical and biochemical systems. Examples of applications range from predicting binding processes of biomolecules for drug design \cite{Williams-Noonan2018, Cournia2017, Christ2014} to determining thermodynamic properties of crystalline materials \cite{Swinburne2018, Freitas2018, deKoning1999}. For large and complex systems with slow relaxation rates and typically $10^5$ to $10^7$ particles, only limited accuracy is achieved \cite{Zuckerman2002}, despite substantial methodological 
progress \cite{Jarzynski1997, Vaikuntanathan2008, Valsson2014, Shirts2003} and immense computational effort. Besides force field inaccuracies, insufficient sampling is the 
main bottleneck \cite{Aldeghi2018}. Here, we develop and evaluate a variational approach for optimal sampling that minimizes the sampling error.

Given the Hamiltonians $H_1(\x)$ and $H_N(\x)$ of two states $1$ and $N$, where $\x\in {\rm I\!R}^{3M}$ denotes the position of all $M$ particles of the simulation system, the free energy difference $\Delta G_{1,N}$ between these states
is given by the Zwanzig formula~\cite{Zwanzig1954},
\begin{align}
	\Delta G_{1,N}= -\ln\langle e^{-[H_N(\x) - H_1(\x)]})\rangle_1\,,
	\label{eq:zwanzig}
\end{align}
where $\langle\rangle_N$ denotes an ensemble average defined by $H_1(\x)$, which is approximated by averaging over a finite sample
of size $n$ obtained from atomistic simulations or Monte Carlo sampling.  For ease of notation, $k_BT= 1$.

\begin{figure}[t]
	\includegraphics[width=8.6cm]{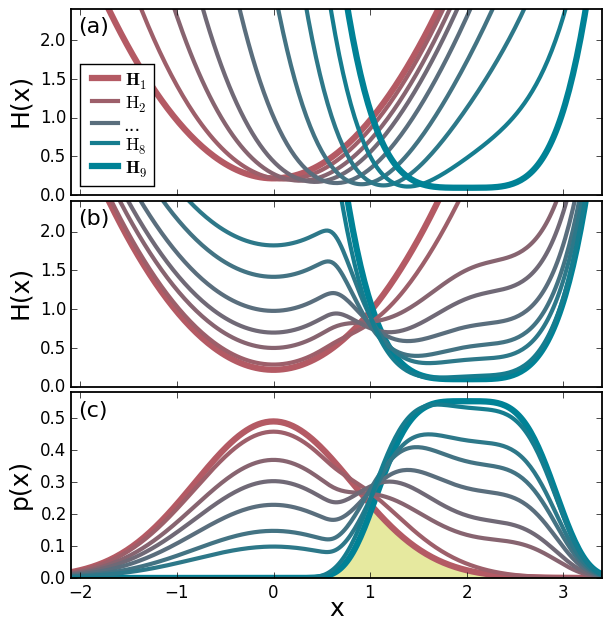}
	\caption{Sequences of intermediates between a harmonic potential $H_1(\x)=\frac{1}{2} x^2+b$ and a quartic potential 
		$H_9(\x)=(x-x_0)^4+c$ (thick lines), where $b$ and $c$ have been determined such that $Z_1 = Z_9 = 1$, i.e., $\Delta G_{1,9} = 0$.
		(a) A linear interpolation between $H_1(\x)$ and $H_9(\x)$. For better visualization, the intermediates were vertically offset to align the minima. (b) Intermediate Hamiltonians and (c) resulting configuration space densities of VMFE. The yellow area highlights the configuration space density overlap $K$ between states $1$ and $9$.}
	\label{fig:01_morphing_paths}
\end{figure}

Alchemical transformations substantially reduce sampling errors \cite{Lu2001a, Lu2001}
by introducing $N-2$ intermediate states $s$,
\begin{align}
	H_s(\x) = (1-\lambda_s)H_1(\x) + \lambda_s H_N(\x), \;\; \lambda_s \in [0, 1], %\;\; s=1\ldots N,
	\label{eq:lin_interpolation}
\end{align} 
and accumulating small free energy differences between all adjacent states $s$ and $s+1$,
\begin{align}
\Delta G_{1,N}= \sum_{s=1}^{N-1}\Delta G_{s,s+1}\,.
\label{eq:fep_sum}
\end{align} 
This technique is also employed in other fields, for example in the context of Bayesian statistics, where the plausibility of two different models is compared by calculating their marginal likelihood ratio~\cite{Gelman1998, Habeck2012}. With few exceptions~\cite{Christ2007,Pham2012}, only the linear interpolation between $H_1$ and $H_N$ of Eq.~(\ref{eq:lin_interpolation}) is used, that is illustrated for a simple one-dimensional case in Fig.~\ref{fig:01_morphing_paths}(a).

Here, we will generalize this linear interpolation for two of the most established methods, the Free Energy Perturbation (FEP)~\cite{Zwanzig1954} and the Bennett Acceptance Ratio (BAR) method~\cite{Bennett1976}. Specifically, we ask which sequence 
$H_2(\x)\ldots H_{N-1}(\x)$ amongst \textit{all possible functionals} $\{H_s[ H_1, H_N]\}$ yields, on average, the highest accuracy.
Figure~\ref{fig:01_morphing_paths}(b) and \ref{fig:01_morphing_paths}(c) show such a general interpolation sequence, which we refer to as {\em Variational 
Morphing Free Energy} (VMFE) method. Unexpectedly, the result will also turn out to be a generalization of BAR to any $n$ and $N$. 

Note that our approach differs from previous attempts, such as soft-core potentials \cite{Steinbrecher2007}, where {\em ad hoc} functionals are used. For linear interpolations (Eq.~(\ref{eq:lin_interpolation})), the distribution of $\lambda$ points has been optimized \cite{Naden2014} which is also not the general solution we aim for.

\begin{figure}[t]
	\includegraphics[width=8.6cm]{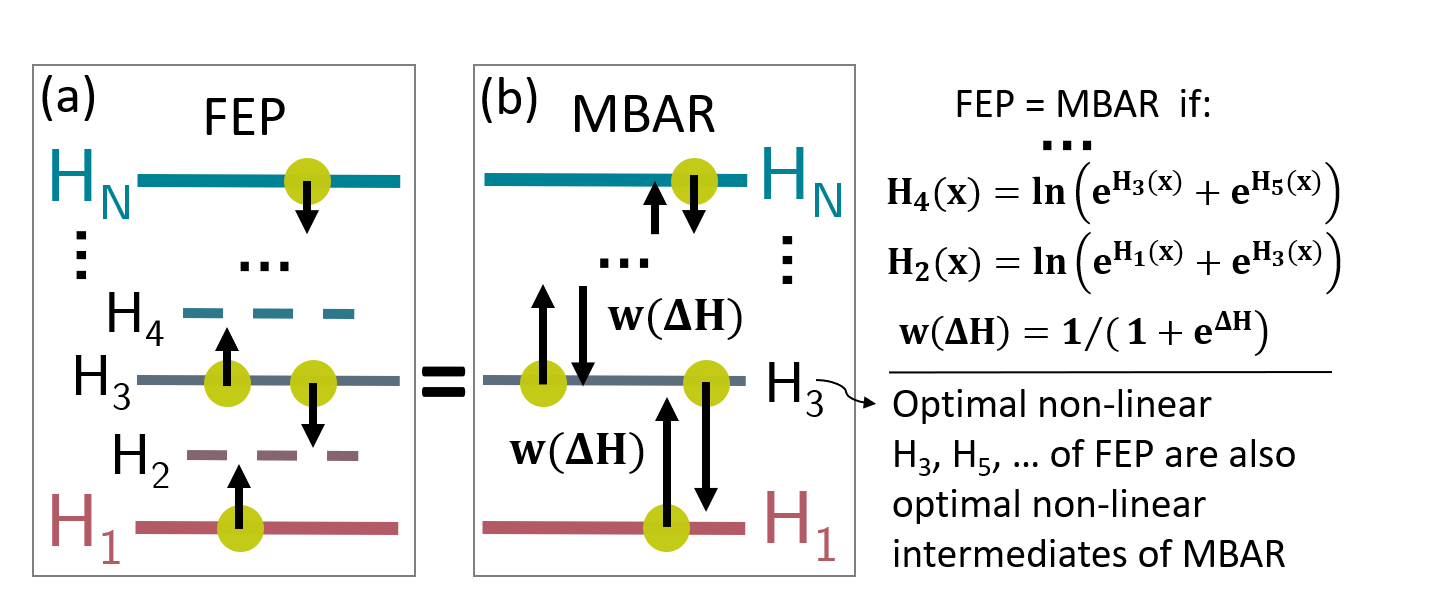}
	\caption{Two schemes of free energy calculation. Yellow dots represent sample sets in the respective potential; arrows indicate the evaluation of differences $\Delta H(\x)$ between adjacent Hamiltonians. Free energy differences are either determined by (a) 
		the Zwanzig formula (FEP), or by (b) BAR with multiple steps (MBAR). Both schemes give identical results at the stated conditions.}
	\label{fig:fep_mbar_illust}
\end{figure}

To solve the above variational problem and to find the optimal sequence of $H_s$, we 
consider the FEP scheme, displayed in Fig.~\ref{fig:fep_mbar_illust}(a), as one possible implementation 
of Eq.~(\ref{eq:fep_sum}) using Eq.~(\ref{eq:zwanzig}). In this particular variant, which is symmetric with respect to exchange of the two end states to
avoid hysteresis effects, sample points are solely drawn from the odd-numbered 'sampling states', and not from the 
even-numbered 'target states'.  The average accuracy of this scheme is the average over all 
sampling realizations of the mean-squared deviation (MSD) 
of the free energy difference $\Delta G_{1,N}^{(n)}$ from the exact difference $\Delta G_{1,N}$,

\begin{equation}
\scalebox{0.92}[1]{$ \sigma ^2 =\:  \EX\left[ \left( \Delta G_{1,N} - \sum_{\substack{s=1 \\ s \; \text{odd}}}^{N-2} \left(\Delta G_{s \rightarrow s+1}^{(n)}  - \Delta G_{s+2 \rightarrow s+1}^{(n)}\right)\right)^2\,\right].$}
\label{eq:acc_multiple}
\end{equation} 
As in Fig.~\ref{fig:fep_mbar_illust}, the arrows point from sampling to target states.

Assuming for each sample state $s$ a set of $n$ independent sample points $\{\x_i\}$, drawn from ${p_s(\x)=e^{-H_s(\x)}\slash Z_s}$, 
with partition function $Z_s$,
the terms arising from expanding Eq.~(\ref{eq:acc_multiple}) will be considered one by one. For the linear term, the average over all sample realizations reads

\begin{equation}
\begin{split}
\label{eq:deltag_discrete}
\EX\left[\Delta G_{s \rightarrow s+1}^{(n)}\right] =& - \int p_s(\x_1)\dx_1 ... \int p_s(\x_n)\dx_n\\& \ln \left[ \frac{1}{n} \sum_{i=1}^{n} e^{-(H_{s+1}(\x_i) - H_s(\x_i))}\right],  
\end{split}
\end{equation}
and for the quadratic term
\begin{equation}
\begin{split}
\label{eq:var_expansion}
\EX\left[\left(\Delta G_{s \rightarrow s+1}^{(n)}\right)^2\right] &= \int p_s(\x_1)\dx_1 ... \int p_s(\x_n)\dx_n \\&\left(\ln \left[ \frac{1}{n} \sum_{i=1}^{n} e^{-(H_{s+1}(\x_i) - H_s(\x_i))}\right]\right)^2.
\end{split}
\end{equation}

Similar expressions are obtained for $\Delta G_{s+2 \rightarrow s+1}^{(n)}$. The exact free energy differences are
\begin{equation}
\Delta G_{s,s+1} = -\ln \int  e^{-(H_{s+1}(\x) - H_s(\x))} p_s(\x) d\x\,.
\end{equation}

For shifted Hamiltonians ${H_s'(\x) = H_s(\x) - C_s}$ and ${H_{s+1}'(\x) = H_{s+1}(\x) - C_{s+1}}\,$, 
Eq.~(\ref{eq:zwanzig}) yields 
\begin{equation}
\Delta G_{s' \rightarrow (s+1)'}^{(n)} =  \Delta G_{s \rightarrow s+1}^{(n)} - C_{s+1} + C_s,
\label{eq:zwanzig_invariance}
\end{equation}
which also holds for $\Delta G_{s', (s+1)'}\,$. Because these offsets cancel out in Eq.~(\ref{eq:acc_multiple}), the 
accuracy $\sigma$ is invariant under any choice of offsets $C_s$ and $C_{s+1}$. 
Choosing $C_s$ and $C_{s+1}$ such that the term in the logarithm of Eqs.~(\ref{eq:deltag_discrete}) and (\ref{eq:var_expansion}) is close to one, and thus all $\Delta G_{s' \rightarrow (s+1)'}^{(n)}$ are small with respect to $k_B T = 1$, first order expansion of the logarithm allows to factorize the integrals, and therefore 
\begin{equation}
\EX\left[ \Delta G^{(n)}_{s' \rightarrow (s+1)'} \right] = \Delta G_{s' ,(s+1)'} \;.
\label{eq:est_eq_corr}
\end{equation}

For the cross terms in Eq.~(\ref{eq:acc_multiple}), note that the estimated free energy differences of the individual steps are based on uncorrelated sample sets, and therefore
\begin{equation}
\begin{split}
\EX\left[ \Delta G^{(n)}_{s' \rightarrow t'} \cdot \Delta G^{(n)}_{u' \rightarrow v'}\right] =&  \EX\left[ \Delta G^{(n)}_{s' \rightarrow t'} \right]\,\EX \left[\Delta G^{(n)}_{u' \rightarrow v'}\right] \\
=& \,\Delta G_{s',  t'} \,\Delta G_{u' , v'} \;,
\end{split}
\end{equation}
for $(s'\rightarrow t') \neq (u' \rightarrow v')$. Using Eq.~(\ref{eq:est_eq_corr}), Eq.~(\ref{eq:var_expansion}) yields
\begin{equation}
\begin{split}
\EX\left[\left(\Delta G^{(n)}_{s' \rightarrow (s+1)'}\right)^2\right] =& \frac{1}{n}\int e^{-2(H_{s+1}'(\x)- H_s'(\x))} p_s(\x)\dx\\ 
&+  f_{s'}(\Delta G_{s',(s+1)'}).
\end{split}
\label{eq:deltag_square}
\end{equation}

Inserting Eqs.~(\ref{eq:est_eq_corr}) and (\ref{eq:deltag_square}) into Eq.~(\ref{eq:acc_multiple}),

\begin{equation}
\begin{split}
\sigma ^2 =&  \sum_{\substack{s=1 \\ s \; \text{even}}}^{N-2} \frac{1}{n}\left(\int p_s(\x)\,\dx\, e^{-2(H'_{s+1}(\x)- H_s'(\x))} \right. \\ & \left.+ \int p_{s+2}(\x)\,\dx\, e^{-2(H'_{s+1}(\x)- H'_{s+2}(\x))} \right.\\  & \left.+ g_{s'}(\Delta G_{s',(s+1)'}, \Delta G_{(s+2)',(s+1)'})\; \right) \,,
\end{split}
\label{eq:msd_squared_terms}
\end{equation}

where $f_{s'}$ and $g_{s'}$ denote expressions that only depend on exact free energy differences and thus are dropped for the optimization below. 

With these expressions, the variational problem can be solved analytically. For the \textit{odd-numbered} states $s$, variation of 
$\sigma^2$, Eq.~(\ref{eq:msd_squared_terms}),
\begin{equation}
\frac{\partial}{\partial H_s(\x)}\left(\sigma^2 + \nu \int (e^{-H_s(\x)} - Z_s) d\x\right) \overset{!}{=} 0
\end{equation}
yields
\begin{equation}
H_{s}(\x)=  -\frac{1}{2} \ln \left(e^{-2(H_{s-1}(\x)-C_{s-1})} +  e^{-2(H_{s+1}(\x)-C_{s+1})}  \right), 
\label{eq:one_int_opt_sampling}
\end{equation}
where $Z_s = \int e^{-H_s(\x)} d\x$ is the (finite) partition sum and $\nu$ is a Lagrange multiplier. %= \Delta G^{A\rightarrow 1} - \Delta G^{2\rightarrow 1}}$. 

Similarly, for the \textit{even-numbered} states, 
\begin{equation}
H_s(\x) = \ln \left(e^{H_{s-1}(\x)-C_{s-1}} +  e^{H_{s+1}(\x)-C_{s+1}}  \right).
\label{eq:one_int_opt_target}
\end{equation}

An additive term $C_s$ in Eqs.~(\ref{eq:one_int_opt_sampling}) and (\ref{eq:one_int_opt_target}) was omitted, as it cancels in ${\Delta G_{s-1\rightarrow s}^{(n)}-\Delta G_{s+1\rightarrow s}^{(n)} }$. The result is a set of equations for all states $s$ for which each Hamiltonian $H_s(\x)$ depends only on the two adjacent states. The initial requirement for small $\Delta G^{(n)}_{s'\rightarrow (s+1)'}$ is fulfilled by setting $C_s = -\ln Z_s\,$, as in this case, all $Z_s'$ are one. Rearranging terms for \textit{odd} $s$,

\begin{align}
	e^{-2H_{s}(\x)}= e^{-2H_{s-1}(\x)}\cdot r_{s-1,s}^{-2} +e^{-2H_{s+1}(\x)}\cdot r_{s+1,s}^{-2} %&& \text{for odd } s
	\label{eq:soe_odd}
\end{align}	
and for \textit{even} $s$,
\begin{equation}
e^{H_{s}(\x)} = e^{H_{s-1}(\x)}\cdot r_{s-1,s} +e^{H_{s+1}(\x)}\cdot r_{s+1,s} %&& \text{for even } s
\label{eq:soe_even}
\end{equation}
with $r_{s,t} = Z_s \slash Z_t$. 
The first main result of this letter is the resulting sequence of Hamiltonians that yields the best accuracy for FEP free energy calculations.

The second main result is that Eq.~(\ref{eq:one_int_opt_target}) serves to generalize the BAR method. The latter follows from
Eq.~(\ref{eq:one_int_opt_target}) for $N=3$ with
one intermediate state: Applied to the two involved free energy differences, the Zwanzig formula yields
\begin{align}
\Delta G_{1,3}^{(n)}=& \Delta G_{1 \rightarrow 2}^{(n)} -\Delta G_{3 \rightarrow 2}^{(n)}\\
=& -\ln\langle e^{-[H_2(\x) - H_1(\x)]})\rangle_1 +\ln\langle e^{-[H_2(\x) - H_3(\x)]})\rangle_3.
\end{align}
Inserting Eq.~(\ref{eq:one_int_opt_target}) as the target state Hamiltonian $H_1(\x)$ yields the BAR formula
\begin{equation}
\begin{split}
e^{-(\Delta G_{1,3} - C)} =& \left< \frac{1}{1+e^{H_3(\x)-H_1(\x) - C}}\right>_1 \\ &\Big/ \left< \frac{1}{1+e^{H_1(\x)-H_3(\x) + C}}\right>_3 ,
\label{eq:bar}
\end{split}
\end{equation}
with $C=C_3 - C_1$.  

Notably, the above derivation yields the more general result that Eq.~(\ref{eq:bar}) provides the most accurate free energy estimate also for finite and small $n$, even down to $n=1$ given sufficient configuration space density overlap between adjacent states, which is fulfilled, for instance, in the limit of many intermediates. 
In contrast, because the derivation by Bennett~\cite{Bennett1976} strictly holds only for infinite sampling, so far $n$ was required to be large, and proper convergence had to be assumed. Further, in the original derivation~\cite{Bennett1976} the error distribution of the free energy estimates had to be assumed to be Gaussian, which in our above result is also not required. 
In the context of the Overlap Sampling method \cite{Lu2004}, it has been shown that an FEP intermediate can be defined that yields 
the weighting function from Bennett's derivation; the above results proof that this intermediate is indeed optimal for the FEP scheme.
 
Further generalizing the BAR result, 
Eqs.~(\ref{eq:soe_odd}) and~(\ref{eq:soe_even}) yield optimal 
VMFE intermediates for any (odd) number $N-2$ of intermediate states,
as illustrated in Fig.~\ref{fig:fep_mbar_illust}: For any two sampling states, using BAR and using FEP with the optimal target state of Eq.~(\ref{eq:soe_even}) is equivalent. Applied recursively, therefore, the $\widetilde{N} = (N+1)\slash 2$ sampling states from any sequence of $N$ FEP-optimal Hamiltonians $\{H_s(\x)\}$
are also optimal for multistate BAR (MBAR) \cite{Shirts2008}, where so far, too, only empirically determined linear interpolations have been used as intermediate states. This result, therefore, is a generalization to MBAR.

Conversely, for the setup of one sampling state between two given target end states $1$ and $3$, with remarkable intuition an empirical potential has been proposed~\cite{Christ2007} in the Envelope Distribution Sampling (EDS) method, which is similar to Eq.~(\ref{eq:one_int_opt_sampling}) except for a factor of two in the exponent. In summary, both BAR/MBAR and EDS are special cases of, or approximations to, our more general variational VMFE result that also requires fewer assumptions. 

\begin{figure}[t]
	\includegraphics[width=8.6cm]{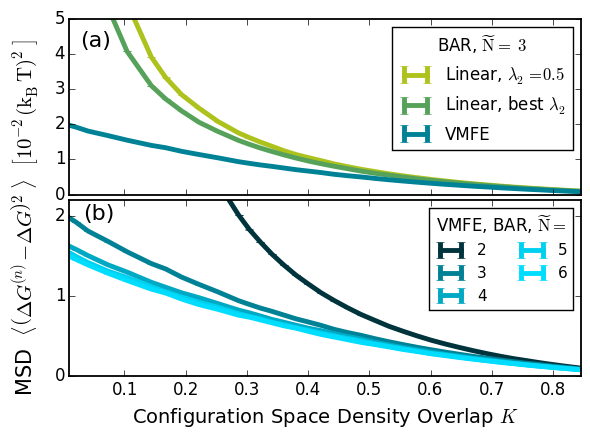}
	\caption{Accuracies of free energy calculations for different overlaps between the end states,
		determined numerically for the model Hamiltonians from Fig.~\ref{fig:01_morphing_paths}. BAR is used between adjacent sampling states.
		(a) Comparison between VMFE and two variants of linear interpolations: a linearly spaced $\lambda_2$ and an empirically optimized $\lambda_2$ yielding the highest accuracy. (b) Accuracies for different numbers of VMFE sampling states for a given total sampling size.}
	\label{fig:results_model}
\end{figure}

To solve Eqs.~(\ref{eq:soe_odd}) and (\ref{eq:soe_even}) for the optimal intermediate Hamiltonians $H_s(\x)$, note that the unknown free energy differences 
$\Delta G_{s,t} = -\ln r_{s,t}$ are part of the equations which, therefore, have to be solved iteratively. With an initial guess for all $r_{s,t}$, the set of equations 
is solved in a point-wise fashion for any given $\x$. After sampling all odd-numbered states, the $r_{s,t}$ values are updated iteratively, such that the sequence of intermediate states converges towards the optimum. For a typical biomolecular many-body system, the additional computational effort is small compared to computing $H_1(\x)$ and $H_N(\x)$. 

For the above illustrative example, Fig.~\ref{fig:01_morphing_paths}(b) and (c) show the optimized Hamiltonians and the configuration space densities, respectively, of the converged sequence of intermediate states. To this end, initial values $r_{s,t}=1$ were used and Eqs.~(\ref{eq:soe_odd}) and (\ref{eq:soe_even}) were iterated until convergence, using numerical integration over $\x$ and updating the $r_{s,t}$ during the process. Unlike the linear interpolations shown in Fig.~\ref{fig:01_morphing_paths}(a), the variational morphing sequence leads to a probability density, which gradually decreases in the region of $A$ and increases in the region of $B$, while remaining almost constant at the point of maximum configuration space overlap. 

Figure~\ref{fig:results_model}(a) shows the results of numerical simulations using the one-dimensional test case shown in Fig.~\ref{fig:01_morphing_paths}. Different minimum distances $x_0$ are used, thereby varying configuration space overlaps $K = \int_{-\infty}^\infty \mathrm{min(p_1(\x), p_N(\x))} \dx$ between the end states, indicated by the yellow area in Fig.~\ref{fig:01_morphing_paths}(b). Sets of $n=100$ uncorrelated sample points are drawn from $p_s(\x)$ through rejection sampling.  $\widetilde{N} =3$ sampling states are used with BAR. For each $K$, the accuracy (Eq.~(\ref{eq:acc_multiple})) is calculated by averaging over 600,000 realizations.  

VMFE (blue curve) yields the smallest MSD for all $K$, compared to both the first linear interpolation variant (light green) using a linearly spaced $\lambda_2=\frac{1}{2}$, like in a typical free energy calculation, and even compared to the second variant (dark green) using the empirically determined $\lambda_2$ value that yields the best accuracy that can be achieved by linear interpolation. For more details, see Supplementary Material. The largest improvements of VMFE are seen for small configuration space density overlaps that notoriously cause the largest uncertainties.

\begin{figure}[t]
	\includegraphics[width=8.6cm]{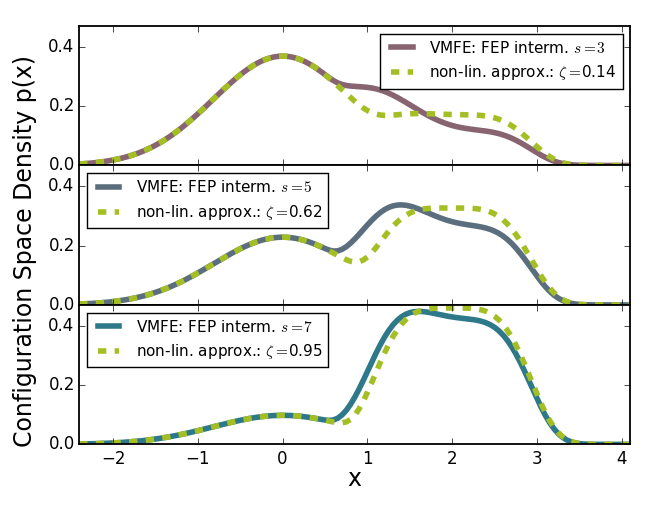}
	\caption{Comparison between configuration space densities of the approximated VMFE sequence (Eq.~(\ref{eq:non_lin_approx}) (dashed lines) 
	with that of the optimal VMFE sequence (solid lines) for the test case shown in Fig.~\ref{fig:01_morphing_paths}(b). 
	For better visualization, the three intermediate sampling states $s=3$, 5, 7 are shown separately.}
	\label{fig:approximation}
\end{figure}  

Figure~\ref{fig:results_model}(b) shows how the accuracy of VMFE improves with increasing number of states $\widetilde{N}$,
keeping the total number of sample points, and hence the total computational effort, constant. For this example, the accuracy increases up to $\widetilde{N}=5$, beyond which no 
further improvement appears.

The above VMFE scheme, Eqs.~(\ref{eq:soe_odd}) and (\ref{eq:soe_even}) couple all intermediates and, therefore, cannot be run in parallel in a straightforward way. This limitation is overcome by two approximations. First, the sampling states are coupled directly using only Eq.~(\ref{eq:soe_odd}). Therefore, while still using BAR between two adjacent sampling states, the corresponding target states are not used for their derivation. 
Second, Eq.~(\ref{eq:soe_odd}) is solved recursively, i.e., the optimal sampling state $H_{\widetilde{N}/2}$ is determined first from $H_1$ and $H_{\widetilde{N}}$, then $H_{\widetilde{N}/4}$ from $H_1$ and $H_{\widetilde{N}/2}$, as well as $H_{3\widetilde{N}/4}$ from $H_{\widetilde{N}/2}$ and $H_{\widetilde{N}}$, and so on. As a result, the approximate intermediate Hamiltonians read
\begin{align}
\scalebox{0.96}{$\hat{H}_s(\x) = -\frac{1}{2}\ln\left[(1-\zeta_s) e^{-2H_1(\x)} + \zeta_s e^{-2(H_{\widetilde{N}}(\x)-C)}\right]$} ,
\label{eq:non_lin_approx}
\end{align} 
with prefactors $\zeta_s$ recursively determined, using Eq.~(\ref{eq:soe_odd}), such that all $\hat{H}_s(\x)$ are a functional of only 
$H_1$ and $H_{\widetilde{N}}$. As above, $C\approx \Delta G$ is determined iteratively. 
Consequently, no prior knowledge of the differences between the individual states is required, and therefore, the sampling simulations for each state can be run in parallel without communication. 

\begin{figure}[t]
	\includegraphics[width=8.6cm]{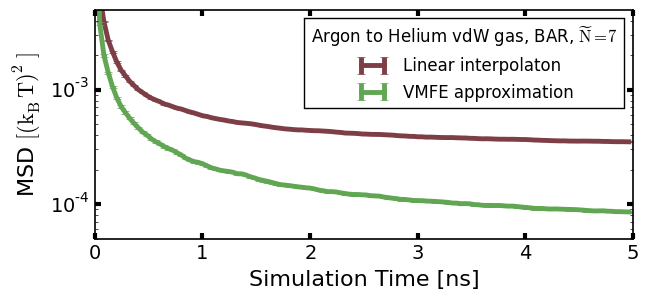}
	\caption{An Argon LJ gas is morphed into a Helium LJ gas. The MSD with respect to a converged reference value is shown depending on the simulation time in each state. Linearly interpolated intermediates (red) and the approximated VMFE sequence (green) are compared with equal spacing of $\lambda$ and $\zeta$ values.}
	\label{fig:vdw_result}
\end{figure} 

Figure~\ref{fig:approximation} shows a comparison between the configuration space densities $p(x)$ of the approximate intermediate Hamiltonians $\tilde{H}_s(\x)$ (dashed lines) with those of the optimal 
${H}_s(\x)$ (solid lines), corresponding to the densities in Fig.~\ref{fig:01_morphing_paths}(c). The two sequences are indeed very similar.
Even if the optimal $\zeta_s$ values are not known {\em a priori}, the approximated VMFE sequence covers the transition behavior of the optimal sequence well, particularly for larger numbers of intermediates.

As a more high-dimensional test case, we calculate the free energy difference between an Argon and a Helium Lennard-Jones (LJ) gas (parameters from \cite{White1999}) with $M=20$ atoms. Fig.~\ref{fig:vdw_result} shows the accuracy, determined through comparison to the result of a converged reference simulation, obtained by approximated VMFE with that of linearly interpolated intermediates. For more details, see Supplementary Material. At 5~ns, an over 4-fold improved accuracy is achieved by VMFE (green) compared to a conventional linear interpolation (red). Conversely, the accuracy achieved by linear interpolation at 5~ns is already obtained at 0.56~ns by VMFE, which thus requires almost 10 times less sampling.

Interestingly, apart from different factors in the exponent, the intermediates of the approximated sequence resemble
those suggested in the context of thermodynamic integration (TI) \cite{Kirkwood1935}. Using approximations to the solution of the
optimization problem for TI for several special cases \cite{Gelman1998}, an expression similar to the approximate 
Eq.~(\ref{eq:non_lin_approx}) was obtained \cite{Pham2012, Blondel2004}. These results require a proper choice of $\lambda$, and it is unclear if the optimal $\lambda$ states are the same for the different methods. Nevertheless, the similarity is striking and suggests 
that our result may also allow further improvements of TI. 

In summary, we derived the optimal accuracy sequence of intermediate Hamiltonians for free energy perturbation calculations. Compared to the
established linear intermediates, the accuracy improvement is substantial, especially for the critical small configuration space density overlap of the end states
that are a hallmark of complex systems. The optimal sequences are fundamentally different from the linear ones, suggesting potential improvement, also for other methods that rely on intermediate states, e.g., TI or non-equilibrium methods \cite{Jarzynski1997, Shirts2003}. 

VMFE was derived assuming statistically independent sampling points $\x_i$. For atomistic simulation based sampling, as well as, to a lesser extent, for MC sampling, subsequent sampling points are correlated, however, particularly when the relevant configuration space densities are separated by large barriers. 
In these cases, when combined with enhanced sampling techniques, such as Hamiltonian replica exchange~\cite{Swendsen1986, Liu2005, Tan2017}, appropriate biasing potentials~\cite{Grubmuller1995,Steiner1998,Laio2002}, or
a combination thereof, VMFE should also yield improved accuracy, albeit the obtained intermediate Hamiltonians will not be optimal due to the 
neglected time correlations. On a more fundamental level, the equivalence of FEP and BAR established here implies that advances in any of these will 
benefit the other.

\appendix

\bibliography{variational_morphing}

\section{Supplementary Material}

\subsubsection{One-dimensional Test Case - Highest Accuracy Linear Interpolation}

Figure~3(a) shows a comparison of the accuracy obtained by VMFE with two variants of a linearly interpolated sequence. As $\widetilde{N} = 3$, sampling is conducted in one intermediate state and the two end states. Sets of $n=100$ sample points are drawn from the corresponding $p_s(\x)$ through rejection sampling, based on which a free energy estimate between the end states is calculated. 

For the linearly interpolated sequence, $\lambda_2$ can be chosen by the user. To empirically obtain the $\lambda_2$ that yields the highest accuracy (dark green), we loop over the allowed range between zero and one in steps of 0.01. To reliably calculate the MSD with respect to the exact value, for each $\lambda_2$ 150,000 free energy estimates are calculated. Once the highest accuracy $\lambda_2$ is determined, the corresponding MSD is calculated once again using 600,000 repetitions. The result of these is shown in the figure. The procedure is repeated for each value of $K$ (42 values). We note that the $\lambda_2$ yielding the highest accuracy varies for different $K$, and is inaccessible in practice for high-dimensional systems.

\subsubsection{Lennard-Jones Gas Simulation}

To compare the accuracy of the free energy estimate using a linearly interpolated sequence of states to the approximated VMFE sequence, a set of free energy calculations between an Argon and a Helium Lennard-Jones gas is conducted.

In each state, $M=20$ atoms are placed at random positions without overlap inside a cubic box. The atoms are assigned velocities drawn from the Boltzmann distribution corresponding to the temperature of $T=298$~K. The simulations are conducted in the NVT ensemble using periodic boundary conditions. The volume of the box is set to (43.5~${\rm \AA})^3$, corresponding to a pressure of about 10~bar. The atomic interaction at a distance $r$ between the centers of two atoms is described through the Lennard-Jones potential,

\begin{equation}
H(r) = 4\epsilon\left[ \left(\frac{\sigma}{r}\right)^{12}- \left(\frac{\sigma}{r}\right)^6\right]
\end{equation}
with parameters $\sigma=3.405$~${\rm \AA}$, $\epsilon=1.0446$~kJ$\slash$mol and $m=39.95$~u for Argon, and $\sigma=2.64$~${\rm \AA}$, $\epsilon=0.0906$~kJ$\slash$mol and $m=4$~u for Helium \cite{White1999}.

At the start, an equilibration run of 1~ns is conducted. The leap-frog algorithm with a time step of 5~fs is used and velocity rescaling at every 20th time step. For both sequences, 800 free energy simulations are conducted with 5~ns simulation time in each state. Five intermediate, i.e., seven states in total are used. In absence of further knowledge, equal spacing of $\lambda_s$ and $\zeta_s$, i.e, $\{0, 0.17, 0.33, 0.5, 0.67, 0.83, 1\}$ is used. For the approximated VMFE sequence, $C=0$ is used throughout the whole simulation. The difference of the Hamiltonians between adjacent states is recorded at every 400th step. Free energy differences are subsequently calculated using BAR. 

A reference free energy difference is determined by conducting a long simulation with each method using 12  states with linearly spaced $\lambda_s$ and $\zeta_s$ values and computation runs of 10~$\mathrm{\mu s}$ in each state. At this length, the relative difference has decreased below $10^{-5}$ ($\Delta G$ = 0.23252 $\mathrm{k_BT}$). Using this reference value, we calculate the MSD of the distribution of 800 free energy differences depending on the simulation time in each state.

\end{document}